# A supreresolution-enhanced spectrometer beyond the Cramer-Rao bound in phase sensitivity

Byoung S. Ham[1,2]
[1]School of Electrical Engineering and Computer Science, Gwangju Institute of Science and Technology, Gwangju 61005, South Korea
[2]Qu-Lidar, Gwangju 61005, South Korea
(Sep. 01, 2024; bham@gist.ac.kr)

**Abstract**
Precision measurement has been an important research area in sensing and metrology. In classical physics, the Fisher information determines the maximum extractable information from statistically unknown signals, based on a joint probability density function of independently and identically distributed random variables. The Cramer-Rao lower bound (CRLB) indicates the minimum error of the Fisher information, generally known as the shot-noise limit. On the other hand, coherence has pushed the resolution limit further overcoming the diffraction limit using many-wave interference strictly confined to the first-order intensity correlation. However, practical implementation is limited by the lithographic constraints in, e.g., optical gratings. Recently, a coherence technique of superresolution has been introduced to overcome the diffraction limit in phase sensitivity using higher-order intensity correlations of a phase-controlled output field from an interferometer. Here, the superresolution is adopted for precision metrology in an optical spectrometer, whose enhanced frequency resolution is linearly proportional to the intensity-product order, overcoming CRLB. Unlike quantum sensing using entangled photons, this technique is purely classical and offers robust performance against environmental noises, benefiting from the interferometer's scanning mode for fringe counting.

To overcome the diffraction limit of classical physics, quantum sensing has been introduced to an interferometer by leveraging the quantum correlation of entangled photon pairs [1-4]. In traditional interferometers such as a Mach-Zehnder or Michelson interferometer, the interference fringes in the output field arise from the first-order intensity correlation [5]. Both coherence and quantum approaches yield the same results for this fringe [6]. However, to fully exploit the advantages of quantum sensing of an unknown signal, interacting photons must be resolved for higher-order intensity correlations [7-13]. To exploit the benefit of quantum sensing, a polarization-projection measurement technique has been developed [8,14], resulting in superresolution of photonic de Broglie waves (PBWs) [7-13]. This projection measurement has been commonly used to demonstrate nonlocal quantum correlation that violates Bell's inequality [15]. Given that quantum techniques rely on the particle nature of photons satisfying a statistical ensemble of events to be independently and identically distributed (*iid*), they must forgo phase information in an interferometer. According to the Heisenberg uncertainty principle, the particle and wave natures must be exclusive in quantum mechanics [16]. Consequently, the phase information of entangled photons must be disregarded unless one relinquishes the particle nature. Therefore, in a phase-controlled interferometer, it is assumed that the entangled photons are temporally coherent with a common phase [16], allowing for the assignment of a relative phase without contradicting their particle nature [17].

Fisher information quantifies the maximum amount of information that can be obtained about an unknown signal based on *iid* random variables (see Fig. 1) [18]. In the context of an interferometer, these random variables can be thought of as polarization bases of light (see Fig. 2). For example, coin tossing represents two random variables: heads and tails. Given a fixed probability for a particular outcome, the likelihood function of coin tossing describes the joint probability of these *iid* random variables. In an interferometer, a single coin tossing corresponds to the first-order intensity correlation, which results in interference fringes. In this case, there is no distinction in the fringes between single photon [19] and continuous-wave (CW) lights [20-22]. However, for the ordered intensity correlation between them, the measurement error depends on the intensity-product order [20], which is equivalent to the joint probability of coin tossing, i.e., the product of the probabilities of individual coin tosses. This modified measurement error in an interferometer represents a variance determined by the Fisher information [18]. Due to the *iid* random variables, a single toss of multiple coins is equivalent to multiple tosses of a single coin [18]. Likewise, multi-photon or CW input can be considered in the same way as in the coin toss if the ordered intensity correlation can be realized [22]. The independent condition for intensity products is satisfied by Poisson-distributed photons [21]. The identical condition is achieved by coherence optics through projection measurements, dividing the output field into multiple segments [20,22]. Recently, the projection measurement technique has been experimentally demonstrated for the Fisher information of the shot-noise limit (SNL) using Poisson-distributed coherent photons [22].

Unlike the demonstration of SNL using non-phase-controlled projection measurements [20,22], the phase of divided output fields from an interferometer can be precisely managed using linear optics, such as a quarter-wave plate (QWP) [23,24]. As shown in Fig. 2, the intensity product of four divided output fields from an interferometer can differ from the SNL case [22], if a QWP is inserted [24]. This phase-controlled intensity correlation has also been demonstrated using coherent single photons for PBW-like superresolution [25,26], where ref. 24 includes both single-photon and CW regimes. In quantum sensing using entangled photon pairs, superresolution is a necessary condition for the Heisenberg limit, although it is not sufficient on its own [27]. With a comprehensive analysis of the coherently excited superresolution beyond Fisher information, this paper focuses more on phase sensitivity of an unknown signal demonstrating an enhanced frequency resolution, surpassing SNL as well as conventional counterparts. To address the scalability challenges posed by complex linear-optics configurations (see Fig. 2) [28], implementing a superresolution-enhanced spectrometer becomes a crucial technical challenge. The key to implement this idea lies in the method of output port's division and its phase control in an individual basis for the intensity product. For this, the fundamental physics of superresolution is briefly overviewed for the phase control of divided *iid* output fields [23,24,28]. Finally, an analytical solution of superresolution in an interferometer is sought for the intensity-product order, exceeding the Cramer-Rao lower bound (CRLB).

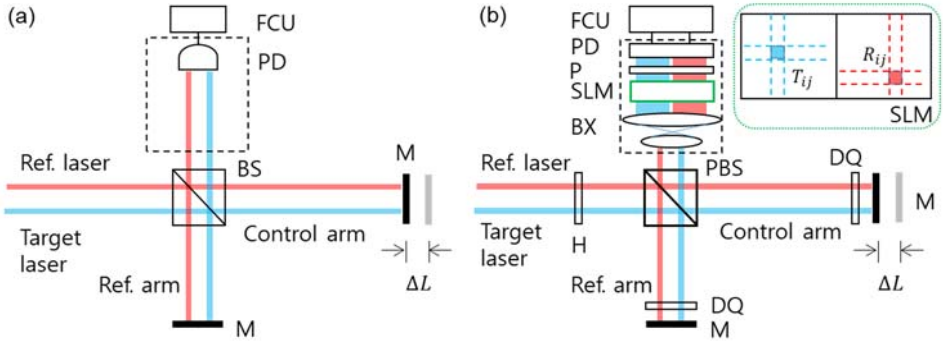

**FIG. 1.** Schematic of superresolution spectrometer. (a) Conventional scheme. (b) Proposed superresolution spectrometer scheme. BS: nonpolarizing beam splitter, BX: beam expander, FCI: fringe counting module, H: half-wave plate, P: polarizer, PD: photodiode, PBS: polarizing BS, DQ: dummy quarter-wave plate, and SLM: spatial light modulator. ΔL is the scan range of the interferometer. $T_{ij}(R_{ij})$: a target (reference) light pixel. The dotted box in (a) represents the SLM block in (b) with no voltage control or P.

Figure 1(a) represents a schematic of a conventional (traditional) spectrometer based on a Michelson interferometer. For the detection of an unknown frequency, the interference fringes are counted for a given continuous path-length scan range ΔL and compared to that of a reference frequency $f_0$. These fringe counts are fairly stable and robust due to the relatively slow phase fluctuation caused by temperatures, mechanical vibrations, and air turbulences, if $\Delta T$ $(= \Delta L/c) <$ μs [29]. For the frequency ($\lambda_T$) measurement of an unknown signal (target laser), the fringe count M is just compared with the known N of the reference ($f_0$), resulting in $f_T = f_0(M/N)$. To work with this scheme, $M = N \pm 1$ becomes the minimum condition for an ultimate resolution $\Delta f$ $(= |f_T - f_0|)$: $(\Delta f)_{min} = f_0 \left|\frac{M-N}{N}\right|$ and $N \geq 2$. Thus, keeping a large N is an essential requirement for better resolution. Due to the stability condition of the interferometer, however, maximum N is upper bound for a given $\Delta T$. This type of a spectrometer has already been widely adopted by modern technologies in academia and industry.

Figure 1(b) represents a schematic of the proposed superresolution-enhanced spectrometer, where the phase-controlled quantum eraser [24,30] is the basic building block, as shown in Fig. 2 [23,24, 28]. To solve the scalability issue, a spatial light modulator (SLM) replaces the linear optics used for the intensity-product ($K$) measurements (see the dotted boxes in Fig. 2 for $K = 4$ [23,24]. As analyzed below, the frequency resolution in Fig. 1(b) is $K$-times enhanced, resulting in $(\Delta f_{SR})_{min} = (\Delta f)_{min}/K$, where $K$ is the pixel number of SLM used for the intensity product. Considering the off-the-shelf million-pixel SLM, a million-folded resolution enhancement is achievable in Fig. 1(b) over the conventional spectrometer in Fig. 1(a).

The theory of the coherently excited quantum eraser [30], as a fundamental element of supreresolution [24,28], is relatively new to general audience. Thus, a brief overview is provided below on how phase control of the output field can achieve superresolution through intensity-product measurements. Given the technological limitations in the scalability of linear optics for the maximum value of $K$ in Fig. 2, the cascaded linear-optics block, which consists of a quarter-wave plate (QWP) and beam splitters (BSs), can be replaced by a SLM as shown in Fig. 1(b). Each voltage-controlled SLM pixel [31] exhibits the same birefringent effect as the QWP [5], enabling independent phase control for all SLM pixels. A proof-of-principle experiment demonstrating macroscopic superresolution using a QWP has recently been conducted for values of K up to 4 [24].

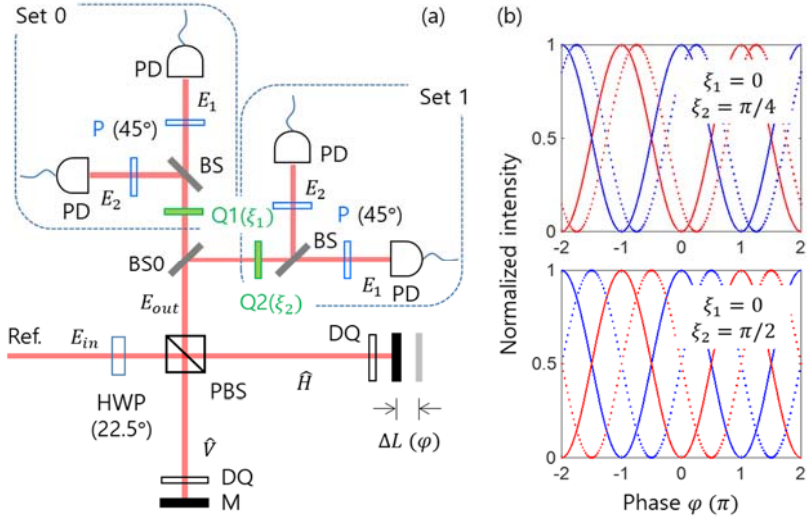

**FIG. 2**. Schematic of superresolution. (a) Schematic of phase-controlled quantum erasers. (b) QWP-dependent fringe shifts. DQ: dummy quarter-wave plate, HWP: half-wave plate, PBS: polarizing beam splitter, BS: nonpolarizaing beam splitter, P: polarizer, PD: photodiode, and Q: quarter-wave plate. $\varphi = 2\pi\Delta L/\lambda$. $\xi_1 = 0$ indicates without Q1. $\xi_1 = \pi/2$ indicates fast-axis vertical in Q2.

To briefly review the phase-controlled quantum erasers, Fig. 2(a) illustrates the basic building block of superresolution depicted in Fig. 1(b), where fringe-shifted quantum erasers enhance resolution in the macroscopic regime (see Fig. 2(b)) [28]. Figure 2 is effective for both reference and target lights in the same way as Fig. 1(b). For the input field $E_{in}$ with statistically random polarization bases ($\hat{H}$ and $\hat{V}$), a half-wave plate (HWP) rotated by 22.5 degrees is inserted to rotate vertically (horizontally) polarized $E_{in}$ to a diagonal (anti-diagonal) direction before entering the Michelson interferometer. Inside the interferometer, a polarizing beam splitter (PBS) establishes a predetermined polarization-path correlation, resulting in no interference fringes at the output port [30]. In the quantum version, this PBS-based Michelson interferometer reveals the particle nature of a single photon in the input light $E_{in}$, leading to distinguishable photon characteristics [32].

In Fig. 2, the output field $E_{out}$ from the modified Michelson interferometer, which includes an HWP and PBS, shows no interference fringes due to the orthogonal polarizations of the light beams [30]:

$$\boldsymbol{E}_{out}(\varphi) = E_{in}(\hat{H}e^{i\varphi} + \hat{V})/\sqrt{2}, \tag{1}$$

where $\hat{H}$ ($\hat{V}$) is the unit vector of horizontally (vertically) polarized light field. The corresponding output intensity is $\varphi$-independent and thus uniform, $I_{out}(\varphi) = I_{in}/2$, where $I_j = E_j E_j^*$.

The quantum eraser is implemented using a polarizer P [30], which is rotated by 45 degrees from the horizontal axis. As a result, the orthogonally polarized lights in Eq. (1) become parallel in a diagonal direction after passing through P. This alignment generates interference fringes, signifying the action of the quantum eraser (see the blue and red curves in Fig. 2(b)) [30]. In the single-photon regime, this phenomenon is recognized as the indistinguishable photon characteristic of the wave nature. Coherently excited quantum erasers have been experimentally demonstrated in both single-photon [30] and CW regimes [24]. For the *Kth*-order intensity product between quantum erasers, the output field $E_{out}$ must be divided into $K$ equal components, as shown in Fig. 2(a). Given the light bandwidth and intensity equality between the divided fields, the *iid* condition is fully satisfied for the measurement events. In each pair of quantum erasers shown in Fig. 2, Eq. (1) is rewritten do describe individual intensities as:

$$I_1 = I_{in}(1 + cos\varphi)/8, \tag{2}$$

$$I_2 = I_{in}(1 - cos\varphi)/8, \tag{3}$$

where the global phase induced to Eq. (3) by BS is omitted, as it does not affect the intensity. Equations (2) and (3) correspond to the blue and red curves in Fig. 2(b). In each set, the paired quantum erasers exhibit an out-of-phase relation, even with phase control by QWP, due to the opposite polarization direction of $\hat{H}$ induced by the BS (see the dotted curves in Fig. 2(b)) [5,23]. Compared to ref. 30, a factor of 2 is applied because PBS directs the entire output field $E_{out}$ into one output port by the inserted dummy QWPs of DQs. Additionally, the *iid* quantum erasers are coherently used for higher-order intensity correlations, with hase control by QWP playing a crucial role [24,28]. It is important to note that the minimum uncertainty in phase estimation for *iid* unknown signals, measured via intensity product without QWPs (Q1 and Q2) and Ps, corresponds to the CRLB for Fig. 1(a) (see below and Fig. 3) [20].

The role of the QWP in each block (set 0 or set 1) in Fig. 2(a) is to create equally shifted fringes, essential for achieving superresolution [23,24,28], which results in the intensity-product-dependent fringe count rate in Fig. 1(b). As analyzed in ref. 28, the general solution for phase-controlled quantum erasers in Fig. 2(a) is given by $\xi_j = \frac{2\pi j}{K}$ for $j = 0, 1, ..., K-1$, where $\xi_j$ is directly related to the QWP's rotation angle for the jth set of the quantum erasers [28]. This general solution for $\xi_j$ has been experimentally validated for values up to $K = 4$ [24]. With the appropriate QWPs, the individual intensities in the jth set in Fig. 2(a) are as follows:

$$I_{1j} = I_{in}[1 + \cos(\varphi - \xi_j)]/8, \tag{4}$$

$$I_{2j} = I_{in}[1 - \cos(\varphi - \xi_j)]/8. \quad (5)$$

Consequently, the normalized *2Kth*-order intensity product across *K* sets of quantum erasers is expressed as [28]:

$$C_{SR}^{(2K)} = \prod_{j=0}^{2K-1} \sin^2(\varphi - \xi_j). \quad (6)$$

Therefore, properly phase-controlled *2K* quantum erasers, as shown in Fig. 2(a), achieve superresolution and can be applied to the voltage-controlled SLM-based spectrometer in Fig. 1(b). As theoretically analyzed, the equivalent and general form of Eq. (6) is $C_{SR}^{(K)} = [1 + \cos(K\varphi)]/2$ [28], where the intensity-product fringes increase linearly by a factor of *K*. Obviously, this quantum effect of superresolution is obtained through a classical coherence-optics method in a macroscopic regime.

The top row of Fig. 3 represents numerical calculations of the fringe-count rate for three different versions of a spectrometer. The bottom row shows the corresponding difference in the fringe-count rates for $\Delta f = f - f_0$ as a function of acquisition time $\Delta T$. Whenever the phases of $I(f)$ and $I(f_0)$ coincide, the intensity difference reaches zero, resulting in a beating phenomenon. Intensities in all panels are normalized for comparison of resolution (fringe counts). The left column represents a traditional spectrometer, as depicted in Fig. 1(a) without the dotted box, with numerical calculations based on Eq. (2) as a function of frequency and scan time ΔT (= ΔL/c). Here, the reference light is denoted by $f_0$, and all other frequencies correspond to the target light being measured. The beating at every 50 $f_0^{-1}$ serves as a reference for the other calculations.

The middle column of Fig. 3 illustrates the intensity product for $K = 10$, corresponding to the left column with the dotted box in Fig. 1(a), representing the general case of Fisher information for SNL [ ]: $C_{SNL}^{(K)} = [(1 + \cos\varphi)/2]^K$. In this setup, the photodetector in Fig. 1(a) is replaced by the SLM block in Fig. 1(b) without voltage control and without P [20]. The numerically calculated phase sensitivity (resolution) improves by a factor of $\sqrt{K}$ near $\varphi = \pm 2n\pi$ (n=0,1,…), as experimentally demonstrated for *K*=1, 2, and 4 (see the Inset) [22]. However, near $\varphi = \pm(2n + 1)\pi$, the resolution deteriorates by a factor of $1/\sqrt{K}$, resulting in no change in the average resolution [20,22]. As shown in the bottom panels, the difference-fringe rate remains unaffected by the intensity-product order *K*, leading to no improvement in the frequency resolution.

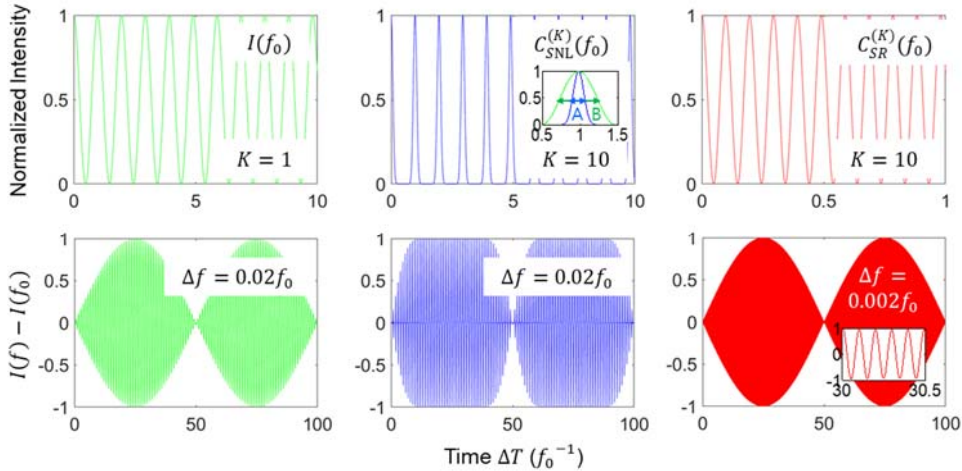

**FIG. 3.** Numerical simulations of the superresolution-enhanced spectrum. (top row) Normalized intensities of K-ordered correlations. (bottom row) fringe difference: $I(f) - I(f_0)$. *I*: Eq. (2), $I_{SNL}$: $K^K$, $I_{SR}$: Eq. (6), where 2K→K. B/A=$\sqrt{10}$.

The right column shows the superresolution effect from Fig. 2 (also shown in Fig. 1(b)), represented by $C_{SR}^{(K)} = [1 + \cos(K\varphi)]/2$. In the upper panel, the fringe-count rate increases tenfold with the intensity-product order $K = 10$. The bottom panel shows that the beating period for the difference-frequency counts shortens by the factor of $K$, resulting in a tenfold improvement in frequency resolution $\Delta f$. This indicates that superresolution enables a $K$-fold enhancement in resolving the frequency of an unknown signal. Consequently, the phase error of an unknown signal with superresolution surpasses the CRLB by a factor $\sqrt{K}$ (shown below): $\Delta\varphi_{SR} = \pi/K$. This represent the Heisenberg limit in phase sensitivity.

The increased fringe-count rate shown in the right column of Fig. 3 can be directly applied to conventional spectrometers. By utilizing a million-pixel SLM block, as depicted in Fig. 1(b), the phase sensitivity of a superresolution-enhanced spectrometer could be improved by up to a million times for an unknown frequency. The fringe counting method for an unknown signal remains effective within the same acquisition time $\Delta T$ for the same scanning mode. Therefore, the environment-free Michelson interferometer greatly benefits the superresolution-enhanced spectrometer, provided that the intensity product can be managed within the same $\Delta T$, i.e., $< \mu s$. Although integrating a million-pixel SLM poses technical challenges, a feasible alternative is to use a 32-channel analog-digital-converter-based multiplexer. This setup allows for a 32-fold enhancement in resolution by incorporating the SLM into a conventional spectrometer.

To analyze superresolution statistically, we examine whether the superresolution-enhanced spectrometer described by Eq. (6) for Fig. 1(b) can surpass the CRLB. For this, the superresolution sample is provided from Eqs. (k4) and (5), resulting in $C_{SR}^{(K)}$ in the right column of Fig. 3, applying the data set $x[n] = A(1 + \cos(\varphi_n')) + w[n]$. Here, $w[n]$ represents white Gaussian noise, A is the intensity of individually measured signals, and $\varphi_n' = \varphi + \xi_n$ denotes the discrete phase control introduced by each pixel of the SLM (or QWP), with $\xi_n = 2\pi n/K$. Thus, the probability density function $p(\mathbf{x}; \varphi)$ is set to calculate Fisher information, where the input laser light satisfies a Poisson distribution with variation $\sigma^2$:

$$p(\mathbf{x}; \varphi) = \frac{1}{\sqrt{2\pi\sigma^2}} \exp\left\{-\frac{1}{2\sigma^2}\sum_{n=0}^{K-1}[x[n] - A(1 + \cos\varphi_n')]^2\right\}. \tag{7}$$

The Poisson distribution is equivalent to the Gaussian distribution if $n \gg 1$. Unlike entangled photon-based PBWs [7-13], which face to issue of a nonvanishing $K = 2$ correlation component leasing to imperfect fringe visibility [14], the visibility of superresolution fringes described by Eq. (6), i.e., $C_{SR}^{(K)}$ in Fig. 3, is nearly perfect for all n, as theoretically [28] and experimentally [24] demonstrated. This is due to the perfect fringe visibility of the related quantum erasers [24,30]. Thus, Eq. (7) is applicable for all n. To determine the minimum variance of the Fisher information, or CRLB, the second derivative of the logarithm of Eq. (7) is calculated as:

$$\frac{\partial^2 \ln p(x;\varphi)}{\partial \varphi^2} = -\frac{A}{\sigma^2}\sum_{n=0}^{K-1}[x[n]\cos\varphi_n' - A(\cos 2\varphi_n' + \cos\varphi_n')]. \tag{8}$$

Upon taking the negative expected value of Eq. (8), we obtain $-E\left[\frac{\partial^2 \ln p(x;\varphi)}{\partial \varphi^2}\right] = \frac{K^2 A^2}{2\sigma^2}$, which is resulted from the phase quantization of the superresolution (see Appendix) [28]. In this context, $K$ coherently prepared identical intensities, each with an equal fringe shift as described in Eqs. (4) and (5), exhibits a similar phase relationship to the equally phase-shifted $K$ amplitudes in a $K$-slit system [5]. Thus, the term $\cos^2 \varphi_n'$ in Eq. (8) results in $K^2/2$ for random phase $\varphi$ (see Appendix). Consequently, the unique feature of the superresolution-based Fisher information yields $\text{Var}(\hat{\varphi}) \geq \frac{2\sigma^2}{K^2 A^2}$, which establishes the corresponding CRLB as $\frac{2\sigma^2}{K^2 A^2}$. Compared to intensity $I$ for $K = 1$ in the left column of Fig. 3, where $\text{Var}(\hat{\varphi}) \geq \frac{2\sigma^2}{KA^2}$ near $\varphi = \pm n\pi$ [18], the superresolution-based spectrometer achives a $\sqrt{K}$ improvement in variance for random $\varphi$. This enhancement in the phase sensitivity matches the Heisenberg limit in quantum sensing [2,3]. Such a phase sensitivity is unattainable with PBW-based quantum sensing unless the nonperfect fringe visibility [27] is addressed [25]. Nevertheless, the macroscopic quantum feature of $\frac{2\sigma^2}{K^2 A^2}$ cannot be obtained by any quantum sensing methods.

In conclusion, we presented and discussed an innovative precision-measurement technique for sensing and metrology that utilizes the intensity product of a phase-controlled output field from an interferometer. This method significantly improved the resolution of an unknown frequency signal, surpassing the conventional Cramer-Rao bound (CRAB) in phase sensitivity. Unlike traditional interferometer-based spectrometers, this superresolution technique achieved a *K*-fold enhancement in frequency resolution. The statistical analysis of the phase-controlled intensity in the macroscopic regime of continuous-wave light demonstrated showing that the new CRLB provides a $\sqrt{K}$ improvement compared to conventional methods of SNL. Importantly, the presented superresolution-enhanced spectrometer retained the same difference-frequency counting measurement technique as traditional spectrometers, ensuring a noise-free scanning mode. Although the intensity-product approach posed technical challenges for high-resolution spatial light modulators due to limitations in analog-to-digital conversion, this superresolution technique holds the potential to revolutionize precision measurements in the future.

This research was supported by the MSIT (Ministry of Science and ICT), Korea, under the ITRC (Information Technology Research Center) support program (IITP 2024-2021-0-01810) supervised by the IITP (Institute for Information & Communications Technology Planning & Evaluation). BSH also acknowledges that this work was supported by a GIST research project grant funded by the GIST in 2024.

**Appendix**

From Eq. (8),

$$\sum_{n=0}^{K-1}[x[n]\cos\varphi'_n - A(\cos 2\varphi'_n + \cos\varphi'_n)],$$

$$= A\sum_{n=0}^{K-1}[(1 + \cos(\varphi'_n) + w[n])\cos\varphi'_n - (\cos 2\varphi'_n + \cos\varphi'_n)],$$

$$= \sum_{n=0}^{K-1}[\cos^2(\varphi'_n) + w[n]\cos\varphi'_n - \cos 2\varphi'_n]. \tag{A1}$$

Taking expectation value of Eq. (A1) is only effective for $\cos^2(\varphi'_n)$ term, whose n-dependent fields are all phase shifted by $\xi_n$. Due to the discretely controlled $\xi_n = 2\pi n/K$, where $\xi_{n+1} - \xi_n = \delta\xi = 2\pi/K$, therefore, the followings are obtained:

$$\sum_{n=0}^{K-1}\cos^2(\varphi'_n),$$

$$= K + 2\sum_{j>i}^{K-1}\sum_{i=0}^{K-1}\cos\varphi'_j \cos\varphi'_i. \tag{A2}$$

For Eq. (A2), $2\sum_{j>i}^{K-1}\sum_{i=0}^{K-1}\cos\varphi'_j \cos\varphi'_i = 2K(K-1)\int \cos^2\delta\xi = K(K-1)$. Thus, $\sum_{n=0}^{K-1}\cos^2(\varphi'_n) = K^2$ is obtained.

**References**

1. C. L. Degen, F. Reinhard, and P. Cappellaro, "Quantum sensing," Rev. Mod. Phys. **89**, 035002 (2017).
2. V. Giovannetti, S. Lloyd, and L. Maccone, "Advances in quantum metrology,". Nat. Photon. **5**, 222-229 (2011).
3. J. P. Dowling, "Quantum optical metrology-the lowdown on high-N00N states," Contemp. Phys. **49**, 125-143 (2008).

4. G. Y. Xiang, B. L. Higgins, D. W. Berry, H. M. Wiseman, and G. J. Pryde, "Entanglement-enhanced measurement of a completely unknown optical phase," Nature Photon. **5**, 43-47 (2011).
5. F. L. Pedrotti, L. M. Pedrotti, and L. S. Pedrotti, *Introduction to Optics*, 3rd ed. (Pearson Education, Inc., New Jersey, 2004), Ch 14.
6. J. Stöhr, "Overcoming the diffraction limit by multi-photon interference: a tutorial," Adv. Opt. & Photon. **11**, 215-313 (2019).
7. J. Jacobson, G. Gjörk, I. Chung, and Y. Yamamato, "Photonic de Broglie waves," Phys. Rev. Lett. **74**, 4835–4838 (1995).
8. K. Edamatsu, R. Shimizu, and T. Itoh, "Measurement of the photonic de Broglie wavelength of entangled photon pairs generated by parametric down-conversion," Phys. Rev. Lett. **89**, 213601 (2002).
9. P. Walther, J.-W. Pan, A. Markus, U. Rupert, S. Gasparoni, and A. Zeilinger, "Broglie wavelength of a non-local four-photon state," Nature **429**, 158–161 (2004).
10. V. Giovannetti, S. Lloyd, and L. Maccone, "Quantum-enhanced measurements: beating the standard quantum limit," Science **306**, 1330–1336 (2004).
11. D. Leibfried, M. D. Barrett, T. Schaetz, J. Britton, *et al*. "Toward Heisenberg-limited spectroscopy with multiparticle entangled states," Science **304**, 1476–1478 (2004).
12. T. Nagata, R. Okamoto, J. L. O'Biran, K. Sasaki, and S. Takeuchi, "Beating the standard quantum limit with four-entangled photons," Science **316**, 726-729 (2007).
13. J. Zhang, M. Um, D. Lv, J.-N. Zhang, L.-M. Duan, and K. Kim, "N00N states of nine quantized vibrations in two radial modes of a trapped ion," Phys. Rev. Lett. **121**, 160502 (2018).
14. F. W. Sun, B. H. Liu, Y. X. Gong, Y. F. Huang, Z. Y. Ou, and G. C. Guo, "Experimental demonstration of phase measurement precision beating standard quantum limit by projection measurement," EPL **82**, 24001 (2008).
15. G. Weihs, T. Jennewein, C. Simon, H. Weinfurter, and A. Zeilinger, "Violation of Bell's inequality under strict Einstein locality conditions," Phys. Rev. Lett. **81**, 5039-5043 (1998).
16. Knight P. & Gerry, C. *Introductory quantum optics* (Cambridge Univ. Press, New York, 2004).
17. D. M. Greenberger, M. A. Horne, and A. Zeilinger, "Multiparticle interferometry and the superposition principle," Phys. Today **46** (80), 22-29 (1993).
18. S. M. Key, *Fundmanentals of statistical signal processing* (Prentice Hall, New Jersey 1993).
19. P. Grangier, G. Roger, and A. Aspect, "Experimental evidence for a photon anticorrelation effect on a beam splitter: A new light on single-photon interferences," Europhys. Lett. **1**, 173–179 (1986).
20. B. S. Ham, "Intensity-product-based optical sensing to beat the diffraction limit in an interferometer," Sensors **24**, 5041 (2024).
21. S. Kim and B. S. Ham, "Revisiting self-interference in Young's double-slit experiments," Sci. Rep. **13**, 977 (2023).
22. S. Kim, J. Stohr, F. Rotermund, and B. S. Ham, "Reducing of the uncertainty product of coherent light through multi-photon interference," arXiv:2404.00496 (2024).
23. B. S. Ham, "Phase-controlled coherent photons for the quantum correlations in a delayed-choice quantum eraser scheme," Sci. Rep. **14**, 1752 (2022).
24. S. Kim and B. S. Ham, "Observations of super-resolution using phase-controlled coherent photons in a delayed-choice quantum eraser scheme," arXiv:2312.03343 (2023).
25. K. J. Resch, K. L. Pregnell, R. Prevedel, A. Gilchrist, G. J. Pryde, J. L. O'Brien, and A. G. White, "Time-reversed and super-resolving phase measurements," Phys. Rev. Lett. **98**, 223601 (2007).
26. C. Kothe, G. Björk, and M. Bourennane, "Arbitrarily high super-resolving phase measurements at telecommunication wavelengths," Phys. Rev. A **81**, 063836 (2010).
27. Nagata, T., Okamoto, R., O'Biran J. L., Sasaki, K. & Takeuchi, S. Beating the standard quantum limit with four-entangled photons. Science **316**, 726-729 (2007).

28. B. S. Ham, "Coherently excited superresolution using intensity product of phase-controlled quantum erasers via polarization-basis projection measurements," Sci. Rep. **14**, 11521 (2024).
29. Kellerer, A. *Assessing time scales of atmospheric turbulence at observatory sites*. Ph.D. thesis (Universit Denis Diderot Paris 7) (2007).
30. S. Kim and B. S. Ham, "Observations of the delayed-choice quantum eraser using coherent photons," Sci. Rep. **13**, 9758 (2023).
31. I. Moreno, J. A. Davis, T. M. Hernandez, D. M. Cottrell, and D. Sand, "Complete polarization control of light from a liquid crystal spatial light modulator," Opt. Exp. **20**, 364-376 (2012).
32. Hardy, L. Source of photons with correlated polarizations and correlated directions. Phys. Lett. A **161**, 326–328 (1992).